\documentclass{aa}
\usepackage{natbib}
\usepackage[utf8]{inputenc}
\usepackage{txfonts}
\usepackage{graphicx}
\usepackage[normalem]{ulem}

\usepackage{amssymb}
\usepackage{amsmath}
\usepackage{color, colortbl}
\definecolor{pad}{rgb}{0.77,0.07,0.77}

\begin{document}

\title{A surprise in the updated list of stellar perturbers of long period comets motion.}

\author{Rita Wysocza\'nska\thanks{\email{rita.wysoczanska@amu.edu.pl}}  and Piotr A. Dybczyński\thanks{\email{dybol@amu.edu.pl}} and Magdalena Poli\'nska\thanks{\email{polinska@amu.edu.pl} }}

\institute{Astronomical Observatory Institute, Faculty of Physics, A. Mickiewicz University, Słoneczna 36, Poznań, Poland 
}

\abstract{The second {\it Gaia} data release ({\it Gaia} DR2) provided us with the precise five-parameter astrometry for 1.3 billion of sources. As stars passing close to the Solar System are thought to be responsible for influencing the dynamical history of long period comets, we update and extend the list of stars that could potentially perturb motion of these comets.}{We aim to announce a publicly available database containing an up to date list of stars and stellar systems potentially perturbing long period comets motion. We add new objects and revise previously published lists. Special emphasis was placed on stellar systems. Discussion on masses estimation is included.}{Using the astrometry, preferably from Gaia DR2, augmented with data from other sources, we calculate spatial positions and velocities for each star. To filter studied objects on the basis of their minimal heliocentric distances we numerically integrate motion of stars under the Galactic potential and their mutual interactions.}{We announce the updated list of stellar perturbers of cometary motion, including masses of perturbers along with the publicly available database interface. These data are ready to be used with the observed long period comets orbits to study an individual influence of a whole sample of perturbers, or specific stars, on a dynamical past or future  of a specific comet. 147 new perturbers were added in comparison to the previously published sources.}
{We demonstrate that a new set of potential perturbers constitute an important tool in studies of cometary dynamics. The usage of our data significantly changes results of the past and future comet motion analysis. We point out a puzzling object in our list, star ALS\,9243. The {\it Gaia} DR2 astrometry suggests a very close encounter of this star with the Sun however, its astrophysical parameters result in a completely different current distance of ALS\,9243 and its large mass.}


\titlerunning{The updated list of stellar perturbers}

\authorrunning{R. Wysocza\'nska et al.}

\maketitle

\section{Introduction}
\label{introduction}
Continuing a longstanding project on obtaining detailed information on the dynamical history of the observed long period comets (LPCs) we have just finished a major update of the potential stellar perturbers list. This was done on the basis of the most recent stellar data, mainly these published as the {\it Gaia} Second Data Release ({\it Gaia} DR2) \citep{Gaia-DR2:2018}.

As it was recently presented \citep{First-stars:2020}, due to a great increase of our knowledge on nearby stars, we are able, in some particular cases, to find stars that can significantly perturb the past (or future) motion of the observed long period comets. Moreover, in this new list of perturbers almost 150 new objects appeared and they significantly change results of dynamical history studies of many LPCs. This research is highly advanced and a detailed paper on LPCs past and future dynamics under the influence of nearby stars and the overall Galactic potential is in preparation.

In this paper we describe in detail the updated version of the list of potential stellar perturbers introduced in the above-mentioned paper. We also announce a simple publicly available database interface to access this list.

It should be stressed that our aim is not to determine and study minimal heliocentric distances of passing stars. We use these nominal minimal distances only as a filtering tool while composing the list. For that reason we do not perform any error budget analysis for these values, as well as for other parameters of the encounters. The uncertainty estimations should be performed individually for each particular star -- comet interaction, for example in a similar way as in \cite{First-stars:2020}. All data necessary to perform such an analysis are included in the announced database.

In Sect.~\ref{compiling_list} we describe sources used while selecting potentially interesting stars. In Sect.~\ref{singlestars} methods applied to single stars and problems concerning estimations of their masses are discussed as perturber mass is crucial in examining mutual interactions between a star and a comet. An in-depth analysis of properties of a single star ALS\,9243, a new, puzzling but potentially strong stellar perturber is presented in Sect.~\ref{als9243}. Sect.~\ref{multiples} and \ref{multiple_examples} focus on multiple systems, the most interesting and troublesome systems are described. In Sect.~\ref{database} a brief description  of a public database where our results are presented is given. 
In Sect.~\ref{usefulness} we include several examples on how the new list of stellar perturbers changes the results of cometary dynamics studies. We also present the dependence of these results on the assumed mass of ALS\,9243. We conclude in Sect.~\ref{conclusions} with a short discussion on main results, issues encountered and prospect for the future.  

\section{Compiling the list of perturbers}
\label{compiling_list}
Using modern data on stellar positions and kinematics we decided to check again all stars mentioned in several published papers on stellar encounters with the Solar System for their minimal distances from the Sun. Our initial list of stars consists of (these sources partially overlap): 

\begin{itemize}
\item[\textbullet] 156 stars (with the proximity threshold (PT) of 5\,pc) from \cite{garcia-sanchez:2001}, based on HIPPARCOS catalogue \citep{hipparcos},
\item[\textbullet] 46 stars (PT=2.5\,pc) listed in \cite{dyb-hab3:2006}, based on ARIHIP catalogue \citep{arihip}, 
\item[\textbullet] 142 stars (PT=5\,pc) analyzed in \cite{jimenez_et_al:2011},  based on HIPPARCOS catalogue \citep{hipparcos}
\item[\textbullet] 90 stars or stellar systems (PT=3.5\,pc) \cite{dyb-kroli:2015}, based on XHIP catalogue \citep{anderson_francis:2011}, 
\item[\textbullet] 40 stars (PT=2\,pc) \cite{dyb-berski:2015}, based on HIP2 catalogue \citep{hip2_book:2007},
\item[\textbullet] 42 stars (PT=2\,pc) found by \cite{bailer-jones:2015}, based on HIPPARCOS \citep{hipparcos}, HIP2 \citep{hip2_book:2007} and Tycho-2 \citep{tycho2-cat:2000} catalogues,
\item[\textbullet] 166 stars (PT=10\,pc) listed by \cite{Bailer-Jones:2018a}, based on {\it Gaia} DR1 \citep{2016A&A...595A...2G} and TGAS \citep{TGAS} catalogues,
\item[\textbullet] 3379 stars for PT=10\,pc listed in \cite{Bailer-Jones:2018}, based on {\it Gaia} DR2 catalogue \citep{Gaia-DR2:2018}, 
\item[\textbullet] 3865 potential stellar perturbers (PT=10\,pc) found by us in a subset of {\it Gaia} DR2 containing over 7 million stars with measured radial velocities,
\item[\textbullet] 3441 objects (PT=10\,pc) selected from all stars with known radial velocity and parallax found in the SIMBAD database in October 2018 (over 2.2 million stars were checked).
\end{itemize}

Stars from the last two sources were pre-selected according to the linear motion approximation. Then we concatenate the last three sources, numerically integrate each star under the Galactic potential, obtain its minimal heliocentric distance and exclude all stars passing farther than 4\,pc from the Sun. This refinement (i.e. applying PT=4\,pc) left, in the combined list of the last three sources mentioned above, only 487 stars selected from {\it Gaia} DR2 and 522 stars selected from the SIMBAD database (these two sets partially overlap). Stars mentioned earlier in at least one of the listed papers, even if new data were adopted and their minimal heliocentric distances increased drastically, were kept for the record. Finally we obtained a list of 820 unique perturbers with 147 new objects which for the first time are identified as potential perturbers.

\subsection{Models and methods}
\label{models-methods}
While we use the minimal Sun -- star distance only as a filtering tool for completing our list of potential LPCs motion perturbers,   
it might be important for the reader to know how we calculate this value. 
For all stars and stellar systems the final, nominal smallest heliocentric distance is obtained by a numerical integration of equations of motion formulated in a rectangular Galactocentric frame. We use a reference frame transformation parameters and the Galactic position and velocity of the Sun described in \cite{dyb-berski:2015}.

The main difference between our approach and the one presented in the mentioned paper is that, where possible, while calculating rectangular coordinates of stars, we adopted the distance estimates presented in \cite{B-J-distances:2018}. This way estimated distances were collected for 742 single stars and 85 components of multiple systems. For the stars not included in \textit{Gaia} DR2 we relied on the formula 3 proposed in \cite{B-J:Estimating2015}. 

 As it concerns a model of the Galactic potential, we used the one proposed by \cite{irrgang_et_al:2013}.

At first, after completing the preliminary list of 820 objects, we numerically integrated pairs: the Sun with a star or stellar system under the Galactic potential to the past or future depending on the radial velocity sign and recorded the minimal Sun -- star distance. After rejecting all perturbers with this distance greater than 4~pc (this leaves 643 objects) we repeated the calculation as a single numerical integration now having 644 interacting bodies (all perturbers plus the Sun).

It should be stressed here, that finding the minimal Sun -- star distance serves here only as an approximate tool for filtering potential perturbers. These perturbers are intended to be used in dedicated studies of cometary dynamics, where a detailed analysis of the uncertainties should be carefully done taking into account both stellar and cometary data errors. An example of such an analysis one can found in \cite{First-stars:2020}.

Readers interested just in Sun -- star encounters and their uncertainties should consult other papers, for example \cite{dyb-berski:2015} and  \cite{Bailer-Jones:2018}.

\section{Single stars}
\label{singlestars}

Single stars were proceeded in a standard way, described in Section \ref{models-methods}. In a great majority of cases we use the astrometry from {\it Gaia} DR2 together with the radial velocity from the same catalogue, if available. When a star is absent in this source we use the SIMBAD and VIZIER databases to find the most appropriate data. Having positions, proper motions, distance estimates, and radial velocities, we calculate rectangular components of the spatial position and velocity.
Since we are collecting potential perturbers of the long period comets motion, we need one more parameter - a mass of the perturber.

\subsection{Masses}
To complete the list of stellar perturbers it was necessary to obtain stellar masses. As {\it Gaia} DR2 do not provide us with masses of stars, we had to search for them in other sources. 

We were unable to find a catalogue or literature source of masses which would cover all stars in question, therefore, our choice was to gather as many different sources of stellar mass estimates as possible, even if, for particular stars, these sources overlap. This approach facilitated a verification of mass estimates and showed whether there is a compliance between different sources and methods. 

Below, we describe specific sources and methods which allowed us to obtain stellar mass estimates with clear indication how many masses were collected with each of these methods. 

405~stars from our list have their mass estimates presented in \cite{Bailer-Jones:2018}. 

572~masses of single stars can be found in \cite{Anders:2019}.
These two above-mentioned sources contain only masses of stars included in {\it Gaia} DR2.

We also performed our own estimations. For M dwarfs we used a formula from \cite{benedict2016solar} which allows us to estimate masses $M$ as a function of the absolute brightness $M_K$
\begin{equation}
    \begin{aligned}
        M&=C_0+C_1(M_K-x_0)+C_2(M_K-x_0)^2+C_3(M_K-x_0)^3\\
        &+C_4(M_K-x_0)^4,
    \end{aligned}
\end{equation}
were polynomial coefficients are $C_0=0.2311$, $C_1=-0.1352$, $C_2=0.0400$, $C_3=0.0038$, $C_4=-0.0032$ and the magnitude offset equals $x_0=7.5$. We use the K-band because it better agrees with the model, as stated in \cite{benedict2016solar}. Using this method we were able to obtain masses for 74 single stars from our list of perturbers.

Masses of main sequence dwarfs with known effective temperatures were estimated by us utilising formulas from \cite{eker2018interrelated}:
\begin{itemize}
    \item for ultra low masses in range $0.179<M/M_\odot\leq0.45$ we use the following formula
    \begin{equation}
        \log(L)=2.028(135)\log(M)-0.976(070)
    \end{equation}
    \item for very low masses in range $0.45<M/M_\odot\leq0.72$ 
    \begin{equation}
        \log(L)=4.572(319)\log(M)-0.102(076)
    \end{equation}
    \item for low masses in range $0.72<M/M_\odot\leq1.05$ 
    \begin{equation}
        \log(L)=5.743(413)\log(M)-0.007(026)
    \end{equation}
    \item for intermediate masses in range $1.05<M/M_\odot\leq2.40$ 
    \begin{equation}
        \log(L)=4.329(087)\log(M)+ 0.010(019)
    \end{equation}
    \item for high masses in range $2.40<M/M_\odot\leq7$ 
    \begin{equation}
        \log(L)=3.967(143)\log(M)+ 0.093(083)
    \end{equation}
    \item for very high masses in range $7<M/M_\odot\leq31$ 
    \begin{equation}
        \log(L)=2.865(155)\log(M)+ 1.105(176).
    \end{equation}
\end{itemize}
For each star its mass was calculated with all the above-mentioned formula and then checked for which formula the estimated mass falls within its range of validity. Thanks to these formulas we managed to obtain 402 stellar masses which were further verified whether they meet the conditions stated in Table~5 from \cite{eker2018interrelated}. After the verification we ended up with 310 masses obtained with this method.

Formulas from \cite{eker2018interrelated} have to be used in conjunction with formulas from \citep{andrae2018gaia} which allow to calculate a radius and a luminosity of the star when only an effective temperature is given.  

Additionally, utilising the luminosity in $J$ band, 473 masses were obtained from the tables created by \cite{Mamajek2013}. Using the same tables 445 masses were gathered basing on the luminosity in $K_s$ band and 495 masses when luminosity in $V$ band was used. Each time, when possible, it was checked whether the effective temperature matches the calculated mass. 

Because {\it Gaia} DR2 provides us with luminosities in $G$ band it was necessary to convert them into other bands. The following formulas from {\it Gaia} DR2 documentation \cite{Gaia-DR2:2018} were used:
\begin{itemize}
    \item to convert into $V$ band
    \begin{equation}
    \begin{aligned}
        G-V&=-0.01760-0.006860(G_{BP}-G_{RP})\\
        &-0.1732(G_{BP}-G_{RP})^2
    \end{aligned}
    \end{equation}
    \item to convert into $J$ band
    \begin{equation}
    \begin{aligned}
        G-J&=-0.01883+1.394(G_{BP}-G_{RP})\\
        &-0.07893(G_{BP}-G_{RP})^2
    \end{aligned}
    \end{equation}
    \item to convert into $K_s$ band
    \begin{equation}
    \begin{aligned}
        G-K_s&=-0.01885+2.092(G_{BP}-G_{RP})\\
        &-0.1345(G_{BP}-G_{RP})^2.
    \end{aligned}
    \end{equation}
\end{itemize}

For 702~single stars their mass estimates were directly obtained from \cite{Mamajek2013} tables using only an effective temperature given. 

The TESS catalogue  \citep{stassun2018tess,muirhead2018catalog} was also used as a source of star mass estimates. From TESS1 325~masses were gathered, while in TESS2 there were no masses of stars in question. 

Using all the above-mentioned sources, for most of the stars, we have obtained several, sometimes different, mass estimates. For a great majority of them (572 from among 783 single stars) we finally decided to use the mass estimates from \cite{Anders:2019}. For objects missing in this source  we decided to take a mean of the gathered values after the most extreme ones and these most flawed have been excluded.

Masses of components of multiple systems were obtained in a similar way depending on the data availability. 76 stellar masses of components of multiple systems recognised by {\it Gaia} DR2 were gathered from \cite{Anders:2019}. We also used tens of other sources found through the SIMBAD and VIZIER databases. In some cases mass estimates were taken from papers that describe the specific multiple system. 

\section{New, puzzling but potentially strong stellar perturber: ALS\,9243}
\label{als9243}

Using the astrometry from the {\it Gaia} DR2 catalogue and the radial velocity from the SIMBAD database we found that a star ALS\,9243, never mentioned in earlier papers in a context of being a cometary motion perturber, 2.5\,Myr ago passed as close as 0.25\,pc from the Sun. But the main reason for being surprised was the estimated mass of this star: according to the spectral type O9 -- B0  and the luminosity class IV repeated in the literature we should assume its mass to be over 15 solar masses! Such a close passage of such a massive star that took place only 2.5 Myr ago would have made a strong influence on the observed long period comet orbits and probably on the Solar System as a whole. At first, we classified this object as a multiple star due to the information from the SIMBAD database but later it appeared that its multiplicity is rather not confirmed. Although this object can be found in the WDS catalogue \citep{Mason2001} basing on observations done by \cite{Aldoretta:2015}, there is no indication of a name of the alleged second component and data necessary to calculate its position and velocity.

\subsection{What we know about the star ALS 9243} \label{sec:what_we_know}

The star in question was probably first mentioned and named in 1965 during the completion of the 'Luminous Stars in the Northern Milky Way' catalogue \citep{Nassau:1965,Hardorp:1965}. The star was designated as LS VI -04 19, which reads: Luminous Stars, volume six, declination zone -04, star number 19. This was an objective prism survey aimed at young stars. They quote OB as the 'estimated spectral type'. Almost forty years later an 'all sky' database of OB stars was collected by \citet{Reed:2003,Reed:2005} and he assigned a new name to this star: ALS\,9243. This name will be used throughout the present work. 

Later on, this star has been also included in large modern catalogues: Tycho-2 \cite[as TYC 4809-2410-1,][]{tycho2-cat:2000}, 2MASS \cite[as J06593022-0448438, ][]{2MASS-cat:2003} and finally {\it Gaia} DR2 \cite[as DR2 3101630187797866112, ][]{Gaia-DR2:2018}.

In an elegant paper by \citet{Graham:1971} the photometric distance to this star was first estimated, as well as its radial velocity. In the last row of his Table I one can find a distance modulus equal 11.9$^{m}$ (which is equivalent to the distance of 2.4 kpc, other distance estimates are presented in Table~\ref{tab:distances}) and $v_{r}=49.5$\,km\,s$^{-1}$. In the same year \citep{Crampton:1971} the star was for the first time associated with the H\,II region and its spectral classification was narrowed down to B0 IV. A year later \cite{Crampton:1972} published radial velocity measurements of the star in question, again using the objective prism, ranging from 26 to 55\,km\,s$^{-1}$ during a ten day interval.

Recently a paper by \citet{Anders:2019} appeared, where distances and astrophysical parameters of large number of {\it Gaia} DR2 stars are recalculated. For ALS\,9243 they obtained a distance of about 93 pc and a mass of only 0.65 M$_\odot$. These results strongly depend on {\it Gaia} DR2 results for this star. However, it should be noted, that the most preferred {\it Gaia} DR2 \citep{Gaia-DR2:2018} astrometry quality indicator, a re-normalised unit weight error (RUWE)\footnote{see: Lindegren, L. 2018, Considerations for the Use of DR2 Astrometry, Tech. rep., available at \url{http://www.rssd.esa.int/doc_fetch.php?id=3757412}} for ALS\,9243 equals 6.97 while the upper RUWE limit for 'good' astrometric solutions is 1.4\footnote{see {\it Gaia} Data Release 2 Documentation (release 1.2), Section 14.1.2}.

\begin{table*}
    \centering
    \caption{ALS\,9243 distance estimates and measurements}
    \label{tab:distances}
    \begin{tabular}{l c c} \hline
         \textbf{distance [pc]} &  \textbf{method}  &   \textbf{ref} \\
         \hline
         2400  & photometric & \citet{Graham:1971} \\
         3300 $\pm$ 800 & photometric & \citet{Avedisova:1984} \\
         31 ( $-21\  +\infty$ ) & trig. parallax & Tycho-1  \citep{hipparcos} \\
         86 ( $-35\  +83$ ) & photometric & \cite{Ammons:2006} \\
         256 & photometric & \citet{Pickles:2010} \\
         2900 & photometric & \citet{Garmany:2015} \\
         3200 & photometric & \citet{Aldoretta:2015} \\
         443 & photometric & \cite{stassun2018tess} \\
         94.7 ( $-3.5\ +3.7$ ) & trig. parallax & {\it Gaia} DR2 \citep{Gaia-DR2:2018} \\
         94.5 ( $-3.3\  +4.0$ ) & trig. parallax & \citet[][based on {\it Gaia} DR2]{B-J-distances:2018} \\
         93 ( $-4\  +6.0$ ) & 'photo-astrometric' & \citet[][based on {\it Gaia} DR2]{Anders:2019} \\
         \hline
    \end{tabular}
\end{table*}

\begin{table*}
    \centering
    \caption{ALS\,9243 spectral classifications}
    \label{tab:spectra}
    \begin{tabular}{l l c c c} \hline
         \textbf{Teff [K]} &  \textbf{Spectrum} & \textbf{Luminosity class} & \textbf{M$_{V}$} & \textbf{ref} \\
         \hline
         -- & OB & -- & -- & \cite{Nassau:1965}
         \\
         -- & -- & -- & $-3.4^{m}$ &  \citet{Graham:1971} \\
         -- & B0 & IV & -- & \citet{Crampton:1971} \\
         -- & O9.5 & V & $-2.29^{m}$ & \citet{Georgelin:1973} \\
         -- & B0.5 & V & -- & \citet{Mayer:1973} \\
         -- & O9.5 & V & -- & \citet{Moffat:1979} \\
         6608 & -- & -- & -- & \cite{Ammons:2006} \\
          -- & F5 & V & -- &  \citet{Pickles:2010} \\
         -- & --   & --& $-3.6^{m}$ & \citet{Garmany:2015} \\
         -- & O9.7 & IV & -- & \citet{Aldoretta:2015} \\
         7773 & -- & V & -- & \cite{stassun2018tess} \\
         6185 & -- & V & -- & \cite{Gaia-DR2:2018} \\
         5461 & -- & -- & -- & \cite{Anders:2019} \\
         \hline
    \end{tabular}
\end{table*}

\subsection{Atmospheric parameters of ALS 9243}

We tried to solve this puzzling inconsistency in ALS\,9243 parameters. In January 2020 on our kind request three spectra with the fiber fed echelle spectrograph ESPERO connected to the 2-m telescope in Rozhen National Astronomical Observatory (\cite{Bonev2017}) with resolving power $R \sim$ 40\,000 and in range from 410 to 950\,nm were obtained. In our (still simplified and approximate) analysis presented below we used only one spectrum observed on 9th January due to its better quality, where measured signal-to-noise ratio was between 30-40.

The atmospheric parameters: an effective temperature $T_{\rm eff}$, a surface gravity  $\log g$ and a projected rotational velocity $v \sin i$ were calculated using the {\sc iSpec} code \citep{B-S2019,B-C2014}. The observed spectrum was compared with a grid of fluxes BSTAR2006 \citep{LH2007} created with TLUSTY model atmospheres and SYNSPEC spectra. We used  stellar atmosphere models which are metal line-blanketed, non-LTE, plane-parallel, and we examine hydrostatic atmospheres. 

\begin{figure}
	\includegraphics[width=\columnwidth]{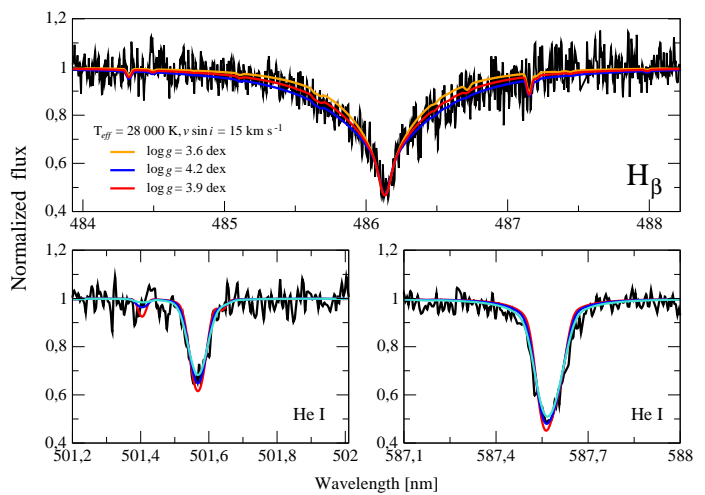}
   \caption{Comparison of the observed spectrum (black color) and synthetic ones (with different colors) of the H$\beta$ region and He I lines. For He I lines different colors correspond to synthetic spectra calculated for various $T_{\rm eff}$, $\log g$  and $v$sin$i$ within the error limits (respectively aquamarine: $T_{\rm eff}$ = 26\,000\,K, logg = 3.7 dex , $v$sin$i$ = 10 km s$^{-1}$ ; blue: $T_{\rm eff}$ = 28\,000\,K, $\log g$ = 3.9 dex, $v$sin$i$ = 15 km s$^{-1}$; red: $T_{\rm eff}$ = 30\,000\,K, $\log g$ = 4.2 dex, $v$sin$i$ = 20 km s$^{-1}$).}
   \label{fig:Lines}
\end{figure}

At first, the effective temperature and the surface gravity were estimated using the Balmer lines H$\alpha$  and H$\beta$. For hot stars ($T_{\rm eff} > 8 000$\,K) Balmer lines are sensitive to the $\log g$ parameter thus both $T_{\rm eff}$ and $\log g$ parameters were derived simultaneously. Additionally, we assumed a microturbulence velocity of 2~\,km\,s$^{-1}$ and a macroturbulence of 0 km s$^{-1}$. The metallicity [M/H] value was fixed to 0.0 dex. In our calculation we also used the six neutral and ionized helium lines (He I, He II), which are well visible in the ALS\,9243 spectrum, such as He II (468.6, 471.4\,nm) and He I (501.6, 587.5, 667.8, 706.7\,nm). The obtained effective temperature is 28\,000 $\pm$ 2\,000\,K, $\log g$ = 3.9 $\pm$ 0.3 and  $v$sin$i$ = 15 $\pm$ 5~\,km\,s$^{-1}$. The comparison of the observed and synthetic spectra within the error limits of H$\alpha$  and two He lines is shown in Fig.~\ref{fig:Lines}.

To evaluate the uncertainties of all determined parameters we took into account the difference in  values calculated separately from the lines. The obtained uncertainties are mainly caused by low signal-to-noise ratio and continuum normalization process of the echelle spectra during which it is difficult to recover the original line profiles. The estimated atmospheric parameters for ALS\,9243 object should be verified in the future from a spectrum of a better quality, with the signal-to-noise ratio of at least 100.

Our temperature measurement is in good agreement with most of the previous spectral type determinations (Table~\ref{tab:spectra}). The obtained temperature and $\log g$ correlate with B0 spectral type and subgiant luminosity class IV. According to \cite{Sterzis} tables this corresponds to the mass of $\sim$ 22 $M_{\odot}$. 

\subsection{Call for observations}

The trigonometric distance obtained by {\it Gaia} DR2 seems to be highly inconsistent with the luminosity and the visual magnitude of such a massive star. The star should be much farther. On the other hand, the J-H color index (fluxes in J and H band were taken from \cite{2MASS-cat:2003}, values of 9.692 and 9.553 respectively, were adopted)  of the star suggests a lower temperature of about 6800\,K. All this contradictory results could be explained for example with the extremely high extinction. In the line of sight we have a molecular cloud SH2-287. However, the cloud might be in the star background or closer to us, and the distance of the SH2-287 is estimated to be 2.1 kpc \citep{Neckel}. The striking discrepancy between the trigonometric and photometric results should be explained with future observations.

We would like to encourage observers to make an effort to clarify this situation by taking high quality positional and spectro-photometric observations of ALS\,9243. Our preliminary attempt revealed that it is a difficult object as it is not so bright and it is visible on the background (or surrounded by) a H\,II nebula. The most striking controversy is the distance: ~0.1 or over 2\,kpc. Also its mass estimates vary from 0.5 up to over 20\,$M_\odot$.

For now, in the absence of a clear explanation of the discrepancies found in the literature, we decided to keep this star in our list, use the {\it Gaia} DR2 astrometry and a 'compromise' mass of 2\,M$_{\odot}$.

\section{Multiples}
\label{multiples}
Almost all previously published lists of stars passing through the close solar neighbourhood (\cite{garcia-sanchez:2001}, \cite{dyb-hab3:2006}, \cite{jimenez_et_al:2011}, \cite{dyb-berski:2015}, \cite{bailer-jones:2015}, \cite{Bailer-Jones:2018}, and  \cite{Bailer-Jones:2018a}) contain only objects considered as single stars even if they are, in reality, parts of multiple systems. Treating components of multiple systems as single stars often leads to misleading conclusions. While a particular component of multiple seems to encounter the Sun at a very small distance, a center of mass of this multiple can even move in another direction. 

Because stellar systems are statistically more massive than single stars, and therefore can act as more significant perturbers, in our work we tried to analyse as many cases of multiplicity as possible. Each star from our list was checked whether it is a component of a system. An identification of multiple systems was done utilizing the SIMBAD database and this explains why our search for multiples was generally limited to stars mentioned in earlier papers. New potential perturbers found only thanks to the {\it Gaia} cannot be checked for multiplicity -- to the best of our knowledge no studies were published  identifying all cases (or even the significant part of cases) of multiple systems in {\it Gaia} DR2.  

While the SIMBAD database facilitated our work on this subject, we had to check each case carefully   by an extended research. Some stars identified in SIMBAD as components of multiple systems are, in fact, single stars, having for example completely different parallaxes. Their alleged multiplicity can, for example, remain from the time where they were observed close to other stars and thought to be dynamically bound to them. On the other hand, although some stars are definitely parts of multiple systems, we sometimes had to treat the whole system as a single star due to the data incompleteness. 

It is important to mention that the most reliable center of mass kinematic parameters can be obtained only when the five-parameter astrometry and the radial velocity are given for all components for the same epoch which is nearly never the case. 

For most of the multiple systems we calculated their center of mass parameters with data available in {\it Gaia} DR2 or with data available in the SIMBAD database. Only in four cases specific systems were described in dedicated papers and we relied on the data found therein. 

In Sect. \ref{multiple_examples} several interesting cases of stellar system investigated by us are described. These systems were either thoroughly examined by us and for the first time classified as multiple stellar perturbers (despite the fact that some individual components of these systems were previously suggested as stellar perturbers) or, by considering their multiplicity, we ruled them out from the list of potential stellar perturbers. Where possible, a comparison with the results found in papers listed in Sect. \ref{compiling_list} is given.

\section{Multiples -- special cases}
\label{multiple_examples}

\subsection{Algol}
Algol, know also as $\beta$ Persei, is a very bright hierarchical system. It consists of a close binary stellar system with a more distant tertiary component. Algol's components were not separated in the SIMBAD database. In the past it was treated as a single star and was classified as stellar perturber by many authors: \cite{garcia-sanchez:2001}, \cite{dyb-hab3:2006}, \cite{jimenez_et_al:2011}, and \cite{dyb-kroli:2015}. Despite a rather distant passage near the Sun (over 3\,pc) this perturber is rather important in near-parabolic comet motion studies due to its large systemic mass and a very small systemic velocity relative to the Sun.

In {\it Gaia} DR2 catalogue one of the components of this system has its right ascension and declination measured but there are no parallax, proper motion, and radial velocity data and, more importantly, there are no clues which component was observed. 

Taking this into account, we decided to rely on data found in the another source.  Based on  observations focused on Algol and UX Arietis, \cite{peterson2011radio} published a parallax, a declination, a right ascension, both proper motion components and a radial velocity of the center of mass of this triple system. We adopted these values which are listed in Table ~\ref{tab:Algol}. 

In view of these values, the Algol system, with its total mass equal to $6.0~ M_\odot$, is the most massive perturber in our list. In fact we use the proximity threshold of 4\,pc when constructing our list of perturbers to keep the Algol system included. Using the input values presented in Table~\ref{tab:Algol} we have obtained a minimal distance of 3.78~pc during the closest approach to the Sun which took place 13.06~Myr ago with the relative velocity  of only 2.17\,km\,s$^{-1}$. Two last values mentioned makes Algol's encounter the oldest and slowest one from among all perturbers in our list. It is worth to mention that the data from  \cite{peterson2011radio} significantly changed these parameters. The previously used values, derived on the basis of \citet{lestrade:1999} or Hipparcos catalogue read: the minimal distance of $\sim3$\,pc, the closest approach at 6 -- 7 Myr ago with the relative velocity of 4\,km\,s$^{-1}$.

\begin{table*}
    \centering
    \caption{Algol system center of mass parameters}
    \label{tab:Algol}
    \begin{tabular}{l c c} \hline
         \textbf{parameter} &  value  &   unit \\
         \hline
         parallax  & 34.7 $\pm$ 0.6 & mas \\
         primary mass & 3.70 & $M_\odot$ \\
         secondary mas & 0.79 & $M_\odot$ \\
         tertiary mass & 1.51 $\pm$ 0.02 & $M_\odot$ \\
         right ascension proper motion & 2.70 $\pm$ 0.07 & mas~yr$^{-1}$ \\
         declination proper motion & -0.80 $\pm$ 0.09 & mas~yr$^{-1}$ \\
         radial velocity & 2.1 & km~s$^{-1}$ \\
         right ascension & $3^h08^m10$\fs13241 $\pm$ 0.7 & mas \\
         declination & 40\degr57\arcmin20\farcs3353 $\pm$ 0.6 & mas \\
         \hline
    \end{tabular}
\end{table*}

\subsection{$\rho$ Orionis}
$\rho$~Orionis is a spectroscopic binary classified as stellar perturber by \cite{dyb-kroli:2015}. It was then treated as a single star, as its components are not listed separately in XHIP \cite{anderson_francis:2011} which was the only source of the 6D stellar data used by the authors. 

Thanks to {\it Gaia} DR2 we were able to update data concerning this system. Both of its components were identified by us in {\it Gaia} DR2 catalogue as {\it Gaia} DR2 3235349837026718976 and {\it Gaia} DR2 3235349940105933568 objects. The second one do not have the radial velocity measured in {\it Gaia} DR2, so we adopted the value from \cite{malaroda2006}.  We used masses from \cite{tokovinin2018multiple} where it is suggested that $\rho$~Orionis is a triple system, but, due to lack of any positional information about the third component, we decided to assume that it is a binary system.

New data allowed us to calculate the center of mass parameters and the new parameters of the approach. While in \cite{dyb-kroli:2015} the minimal distance from the Sun was equal to 3.23~pc, now it is over 17~pc. The encounter happened 2.60 Myr ago at the relative velocity of 46.14~\,km\,s$^{-1}$. 

As it can be seen, an improvement of the quality of the data and the confirmation  of the binary character of the object in question cancelled the importance of $\rho$~Orionis as a stellar perturber of cometary motion.

\subsection{Ross~614}
Ross~614 was at first discovered as a single star by \cite{Ross1927}. Then, \cite{Reuyl1936} detected the second component of this very low mass system which consists of red dwarfs. Later it was a subject of extensive studies, the most recent ones were conducted by \cite{Segransan2000}, \cite{Gatewood2003}, and  \cite{kervella2019stellar}.

Only one component of Ross~614 can be found in the {\it Gaia} DR2 catalogue, where it is identified as {\it Gaia} DR2 3117120863523946368, but it does not have its radial velocity measured. For the second component, even in the SIMBAD database, data are incomplete.

Although above-mentioned papers in-depth examine nature of this stellar system, data found therein are not sufficient enough for our purpose.  

For this reason, our decision was to take masses of both components from \cite{Anders:2019} but use astrometry done only for Ross~614A. Incomplete data from {\it Gaia} DR2 were augmented with radial velocity from \cite{Gontcharov:2006}.

With these values adopted our calculations show that this system encountered the Sun vicinity 0.09 Myr ago at a distance of 3.25~pc. Relative velocity at the time of approach was 27.18~\,km\,s$^{-1}$.

In comparison with results obtained in the past, minimal heliocentric distance of Ross~614 increased. In XHIP catalogue  \cite{anderson_francis:2011}, so also in \cite{dyb-kroli:2015}, it was equal to 3.03~pc. In order to obtain more reliable parameters of the approach new astrometry for the second component (Ross~614B) is needed. 

\subsection{$\alpha$ Canis Majoris}
$\alpha$~Canis~Majoris, known also as Sirius is a visual binary containing Sirius A which is the brightest star in the sky and Sirius B, the nearest white dwarf. There was a long-lasting discussion whether there is a third body in that system because of irregularities in orbits of Sirius A and B.
This possibility was probably ruled out by extensive studies, see for example \cite{Bond2017}.

Sirius was earlier classified as stellar perturber by \cite{garcia-sanchez:2001}, \cite{dyb-hab3:2006}, \cite{jimenez_et_al:2011}, and \cite{dyb-kroli:2015} but in none of these papers the multiplicity was considered.

In {\it Gaia} DR2 only one component is included as {\it Gaia} DR2 2947050466531873024, but it does not have radial velocity measured and there is also no radial velocity in SIMBAD database. For that reason, because we were unable to calculate center of mass parameters, we decided to use values found in \cite{Gatewood1978} augmented with new measurements of masses from \cite{Bond2017}. The values adopted here are listed in Table~\ref{tab:Sirius} were positions and proper motions are given in relation to 1950.0 epoch in FK4 frame. They were further recalculated to be consistent with other data. 

\begin{table*}
    \centering
    \caption{$\alpha$ Canis Majoris center of mass parameters (position and proper motions for 1950.0 epoch in the FK4 frame).}
    \label{tab:Sirius}
    \begin{tabular}{l c c} \hline
         \textbf{parameter} &  value  &   unit \\
         \hline
         parallax  & 0.3777 $\pm$ 0.0031 & mas \\
         primary mass & 2.063 $\pm$ 0.023 & $M_\odot$ \\
         secondary mas & 1.018 $\pm$ 0.011 & $M_\odot$ \\
         right ascension proper motion & -0.0379 & s~yr$^{-1}$ \\
         declination proper motion & -1.211 & ''~yr$^{-1}$ \\
         radial velocity & -7.6 & km~s$^{-1}$ \\
         right ascension & $6^h42^m56$\fs73 & \\
         declination & -16\degr38\arcmin46\farcs4 &  \\
         \hline
    \end{tabular}
\end{table*}

From these data we obtained parameters of the encounter with the Sun which will happen at 2.41~pc in 0.06~Myr. This result is generally in agreement with minimal heliocentric distances obtained earlier. The relative velocity at the time of the approach will be equal to 18.49~\,km\,s$^{-1}$ which makes it a relatively slow passage. $\alpha$~Canis~Majoris system is one of the more massive objects on our perturber list. 

\subsection{$\gamma$ Leonis}

The WDS catalogue (\cite{Mason2001}) identifies four components of $\gamma$~Leonis system. We conducted in-depth investigation to verify whether these components belong to the system. 

While two of them (WDS J10200+1950A and WDS J10200+1950B) have exactly the same parallax of 25.96~mas, for the third one (WDS J10200+1950Ca,Cb) the parallax equals 201.3683~mas, and for the forth one (WDS J10200+1950D) it is measured to be 1.4566~mas. All these values were taken form the SIMBAD database. As it can be seen, only the  first two components actually create a binary system. In the light of available data, two later components were treated by us as single stars that happen to be visually close to the system.  

We decided to calculate the center of mass parameters for $\gamma^1$~Leonis and $\gamma^2$~Leonis. None of the components is in {\it Gaia} DR2 catalogue, so all data were taken from the SIMBAD database. $\gamma^1$~Leonis has its mass (1.41 $M_\odot$) estimated in \cite{Niedzielski2016}. For $\gamma^2$~Leonis we have to adopt a crude mass estimate of 1.50 $M_\odot$ basing on a spectral type. For this system we calculated the minimal heliocentric distance and it came to be 33.32~pc and will occur in 0.26 Myr at the relative velocity equal to 73.06~\,km\,s$^{-1}$. These values ruled $\gamma$~Leonis out from our final list of stellar perturbers.

WDS J10200+1950Ca,Cb is treated as a single star and as such it is included in the RECONS\footnote{http://www.recons.org/} list of the one hundred nearest star systems. It is also included in {\it Gaia} DR2 catalogue as {\it Gaia} DR2 625453654702751872 were it does not have the radial velocity measured. We augmented data from {\it Gaia} DR2 with a radial velocity from SIMBAD database and a mass of 0.467 $M_\odot$ from TESS1 catalogue. Our results show that this star encountered the Sun 0.21 Myr ago at the minimal distance of 3.41~pc and the relative velocity of 17.14~\,km\,s$^{-1}$.

For WDS J10200+1950D, identified also as {\it Gaia} DR2 625453856566097024, data from {\it Gaia} DR2 were used and 1.04 $M_\odot$ mass from \cite{Deka-Szymankiewicz2018} was adopted. The minimal distance of 638.01~pc was obtained and we can state that this star is definitely not a stellar perturber of long period comets motion. 

These examples show that, when it comes to multiple systems, we can not even rely on data found in databases concerning multiple systems  and a careful investigation of each alleged component is necessary.

\subsection{$\alpha$ Centauri}
$\alpha$~Centauri system with its three components is the nearest stellar system to the Sun. It comprise of $\alpha$~Cen~A, a solar-like star, $\alpha$~Cen~B which is a cooler dwarf, and Proxima, a cool red dwarf, recently recognized to be a host of the nearest exoplanet -- Proxima Centauri b.

All of its components were classified as stellar perturbers in \cite{garcia-sanchez:2001}, \cite{dyb-hab3:2006}, \cite{jimenez_et_al:2011}, \cite{dyb-kroli:2015}, and \cite{bailer-jones:2015}, but only in \cite{dyb-kroli:2015} $\alpha$~Centauri was treated as a multiple system and center of mass parameters were calculated to obtain the minimal heliocentric distance. 

Since $\alpha$~Cen~A and $\alpha$~Cen~B are not included in {\it Gaia} DR2 catalogue and Proxima does not have its radial velocity measured in {\it Gaia} DR2 we based our calculation on the heliocentric coordinates given in the Galactic frame found in \cite{Kervella2017Centauri}, summarized in Table~\ref{tab:Centauri}.

\begin{table*}
    \centering
    \caption{Heliocentric coordinates and space velocity components of $\alpha$ Centauri AB and Proxima in the Galactic frame }
    \label{tab:Centauri}
    \begin{tabular}{l c c} \hline
         \textbf{parameter}(unit) &  $\alpha$Cen  &   Proxima \\
         \hline
         X (pc)  & 0.95845 $\pm$ 0.00078 & 0.90223 $\pm$ 0.00043 \\
         Y (pc) & -0.93402 $\pm$ 0.00076 & 0.93599 $\pm$ 0.00045 \\
         Z (pc) & -0.01601 $\pm$ 0.00001 & -0.04386 $\pm$ 0.00002 \\
         XV (kms$^{-1}$) & -29.291 $\pm$ 0.026 & -29.390 $\pm$ 0.027 \\
         YV (kms$^{-1}$) & 1.710 $\pm$ 0.020 & 1.883 $\pm$ 0.018 \\
         ZV (kms$^{-1}$) & 13.589 $\pm$ 0.013 & 13.777 $\pm$ 0.009 \\
         mass (M$_\odot$) & 2.0429 $\pm$ 0.0072 & 0.1221 $\pm$ 0.0022\\
         \hline
    \end{tabular}
\end{table*}

From these values we obtained the center of mass parameters for the system and the parameters of the closest approach to the Sun which will happen in 0.03 Myr at the distance of 0.97~pc, and with the relative velocity of 32.35~\,km\,s$^{-1}$. This result is generally in agreement with the values previously published for components of $\alpha$~Centauri.

\subsection{HD 239960}
HD~239960, known also as Kruger\,60, is a visual binary comprising of two M spectral type stars and it is supposed to be a host of a planetary system (see for example \cite{bonavita:2016}). This stellar system was earlier identified as stellar perturber but its multiplicity has never been taken into account.

Recently both components of Kruger 60 were observed  by  {\it Gaia} and a new astrometry is  available  in the  {\it Gaia}  DR2  catalogue. Components of the system are designated as {\it Gaia} DR2 2007876324466455424 and {\it Gaia} DR2 2007876324472098432. For both of them the radial velocity is missing in this source.

Values from {\it Gaia} DR2 catalogue can be augmented with radial velocities found in the SIMBAD database and other sources. SIMBAD database contain values from the General Catalogue of Stellar Radial Velocities (\cite{wilson:1953}), -24.0~\,km\,s$^{-1}$ for HD\,239960A and -28.0~\,km\,s$^{-1}$ for HD\,239960B, error of radial velocity is in both cases estimated to be 5~\,km\,s$^{-1}$. There is no indication on epochs of observation when these values were measured.  
In the literature it is possible to find values of radial velocities of these stars ranging from -36.0 to -16~\,km\,s$^{-1}$ but often without any information on which component was observed. 

Because the orbital period of Kurger 60 is estimated to be only 44.6 years \citep{bonavita:2016} we aimed to use positions, proper motions, and radial velocities referred to the same epoch. As it was at first impossible, we decided to use available data and calculate position and velocity of the center of mass. Positions, proper motions, parallaxes from {\it Gaia} DR2 were used in conjunction with radial velocities from \cite{wilson:1953} and masses found in \cite{bonavita:2016}. 

While working on the list of stellar perturbers described herein a second interstellar comet 2I/Brisov was discovered. We were involved in a study on its origin and Kruger\,60 seemed to be a good candidate (for more details see \cite{dybczynski:2019}). An in-depth investigation on data available for this stellar system showed us that they are insufficient to obtain reliable results. Thanks to Fabo Feng (via private communication) we have been given an access to the new, unpublished right ascension, declination, parallax, proper motions and the radial velocity of the center of mass of Kruger\,60 which were calculated using PEXO package \citep{FaboFeng:2019} basing on data from HIPPARCOS \citep{hip2_book:2007}, WDS catalogue \citep{WDS:2001} and the recent LCES HIRES/Keck survey \citep{Butler:2017}.

These data were further used to obtain parameters of the approach to the Sun. Kruger\,60 is a future perturber. It will reach its minimal heliocentric distance of 1.81~pc in 0.09 Myr with the relative velocity equal to 38.03~\,km\,s$^{-1}$, This is a slightly smaller distance than presented in the recent paper by \cite{bailer-jones:2015} were Kruger\,60 was classified as a close approaching star but its multiplicity was not considered.

\begin{figure}
    \centering
    \includegraphics[width=\linewidth]{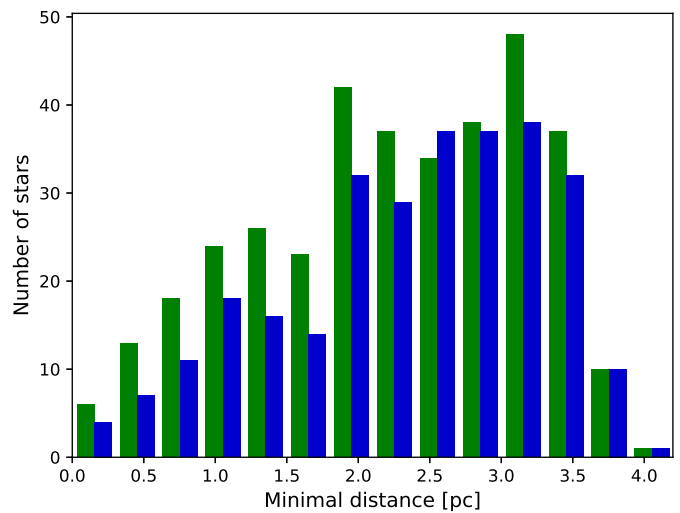}
    \caption{Histogram of minimal heliocentric distances. Distances were divided into two subsets. Green bars correspond to the encounters from the past, blue ones depict the minimal distances of approaches which will occur in the future. 643 stars with the minimal distances smaller than 4.0~pc were considered.}
    \label{fig:hist}
\end{figure}

\begin{figure*}
    \centering
    \includegraphics[scale=1.1]{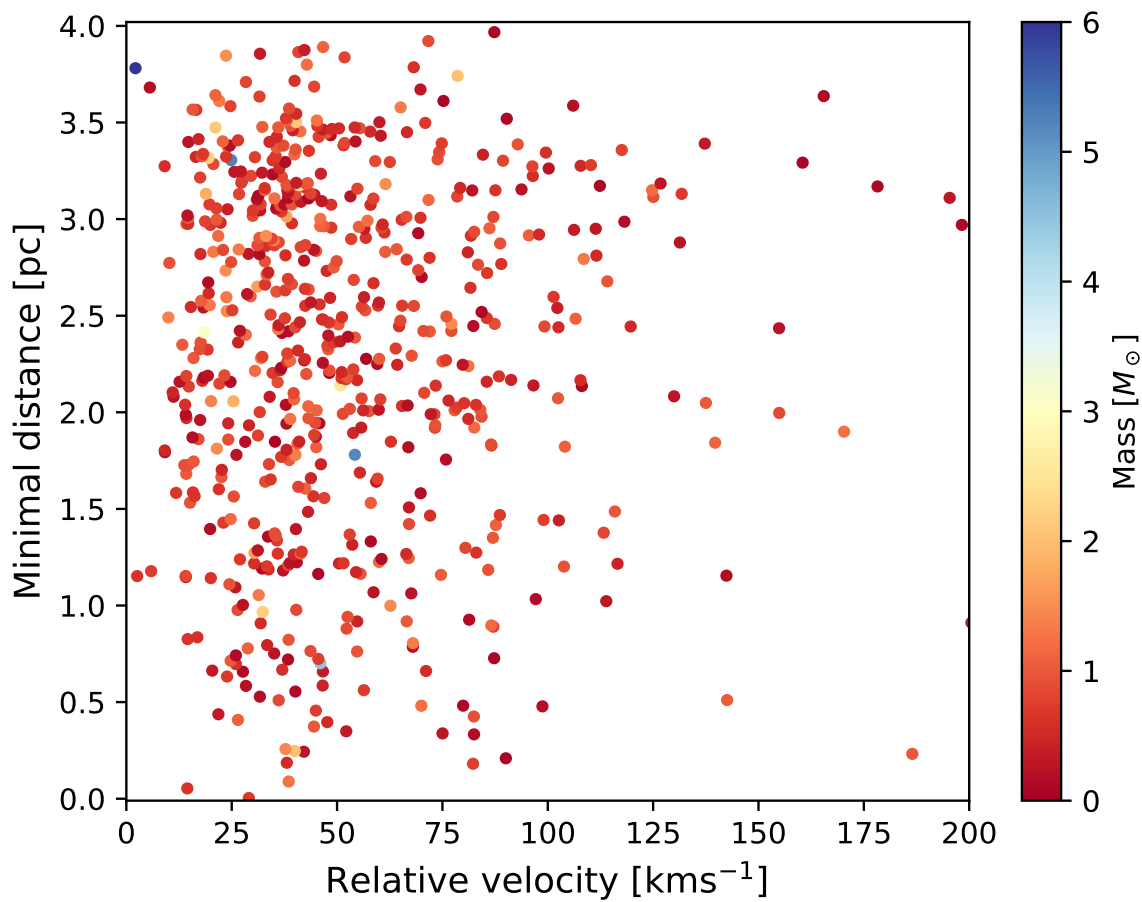}
    \caption{Nominal relative velocities, minimal heliocentric distances, and masses of objects included in our list. 613 stars  were plotted. See text for a explanation. }
    \label{fig:vrelmindist}
\end{figure*}

\section{On-line database for stellar perturbers}
\label{database}

After collecting all necessary data we prepared a simple database containing all the data with their uncertainties and sources. We also included heliocentric rectangular position and velocity components in the Galactic frame together with the adopted mass estimates of all 820 considered perturbers. The whole set of these objects was numerically integrated back and forth in time taking into account all their mutual interactions and including the Galactic overall potential, as described in \cite{dyb-berski:2015}. Our results reveal that 714 of our perturbers encountered or will encounter the Sun within a distance smaller than 10\,pc and 643 were or will be closer than 4.0\,pc. We finally accepted this later proximity threshold  as the one defining the potential perturber, just to keep the Algol system in our list. This system is important because of its large mass of 6\,$M_{\odot}$ and extremely small relative velocity of 2\,km s$^{-1}$. Finally, we keep data for all 820 objects in our database but we name only 643 of them as 'potential perturbers' of the near-parabolic comet motion. It is worth to mention that the final list includes 147 new objects for the first time qualified as potential stellar perturbers of long period comets motion. 

The distribution of the minimal distances between these stars and the Sun is presented in Fig.~\ref{fig:hist}.
The number of the past perturbers is slightly greater than the number of the future ones due to a disproportion in data from the SIMBAD database (there are many more objects with positive radial velocity there).  Such a disproportion can also be noticed in data from the \textit{Gaia} mission, see for example results in \cite{Bailer-Jones:2018} when the sample is limited to stars with minimal heliocentric distance smaller than 4\,pc.

The obtained minimal distances from the Sun are also included in the database for all objects. Objects with the small minimal heliocentric distance will be used by us in the long period comets dynamical studies. All the remaining objects are placed in our database just for the record since they were mentioned in earlier papers. Some of them might return to the list of potential perturbers when new data will be gathered. As a consequence, one can trace how an importance of a specific object changes due to the improvement of data quality. 

In Fig.~\ref{fig:vrelmindist} we also present a three-parameter statistic of stars close passages near the Sun which includes the most important nominal parameters from the point of the star -- comet interaction: minimal distances, relative velocities and masses of the perturbers. To increase the readability of this plot we additionally restrict ourselves to stars passing the Sun with a relative velocity smaller than 200~\,km\,s$^{-1}$. The purpose of this figure is only to graphically illustrate the content of our database, i.e. nominal parameters of potential perturbers. Readers interested in a detailed distribution of stars approaching the Sun and in the completeness of our current knowledge in this field should consult an extensive study by \cite{Bailer-Jones:2018a}. 

Our database is publicly available at the address: https://pad2.astro.amu.edu.pl/pub with a simple interface to access the data and their uncertainties, and the sources. Various lists and statistics are also available and crucial results are made available for the download.

\section{Usefulness and importance of the new list of stellar perturbers}\label{usefulness}

As it was already stated, the aim of collecting the list of stars and stellar systems announced in this paper is to provide a tool for studying past and future motion of long period comets outside the planetary zone. We carefully selected all potential stellar perturbers of the LPCs motion basing on the contemporary sources of stellar data, mainly the \textit{Gaia} DR2 catalogue \citep{Gaia-DR2:2018}. The \textit{Gaia} mission results are revolutionary in this respect -- for example, before this mission we had no more than 130 thousand stellar parallaxes at our disposal, most of them from the Hipparcos catalogue \citep{hip2_book:2007}. Now we can use over 1.3 billion parallaxes \citep[see for example][]{B-J-distances:2018} in our search for potential LPCs motion perturbers.

\begin{table*}
\centering
    \caption{Examples of the LPCs dynamical history results obtained with different models. For four representative comets we present the perihelion distance values: of the original orbit (the first row), at a previous perihelion but calculated without stellar perturbations (the second row), at a previous perihelion as obtained in \cite{kroli_dyb:2017} (the third row) and in the last row also previous perihelion but obtained with the new list of perturbers. In the last three rows a previous perihelion distance is presented in the form of three deciles, 10, 50, and 90 percent.}  
    \label{tab:q-prev}
\begin{tabular}{llllllllllllllll}
\hline
\multicolumn{1}{l}{}  &                           &                           &                           &  &                           &                           &                           &  &                           &                           &                           &  &                           &                           &                           \\
                      & \multicolumn{3}{c}{C/1993 F1}                                                     &  & \multicolumn{3}{c}{C/1997 BA6}                                                    &  & \multicolumn{3}{c}{C/1999 N4}                                                     &  & \multicolumn{3}{c}{C/2006 E1}                                                     \\
\multicolumn{1}{l}{}  &                           &                           &                           &  &                           &                           &                           &  &                           &                           &                           &  &                           &                           &                           \\ \hline
\multicolumn{1}{l}{}  &                           &                           &                           &  &                           &                           &                           &  &                           &                           &                           &  &                           &                           &                           \\
q$_{original}$            & \multicolumn{3}{c}{5.89950 +/- 0.00007}                                           &  & \multicolumn{3}{c}{3.440371 +/- 0.000006}                                         &  & \multicolumn{3}{c}{5.504739 +/- 0.000006}                                         &  & \multicolumn{3}{c}{6.03608 +/- 0.00001}                                           \\
q$_{prev}$ (Galaxy only)  & \multicolumn{1}{r}{7.33}  & \multicolumn{1}{r}{8.16}  & \multicolumn{1}{r}{9.76}  &  & \multicolumn{1}{r}{15.90} & \multicolumn{1}{r}{19.54} & \multicolumn{1}{r}{24.74} &  & \multicolumn{1}{r}{6.36}  & \multicolumn{1}{r}{6.44}  & \multicolumn{1}{r}{6.54}  &  & \multicolumn{1}{r}{22.78} & \multicolumn{1}{r}{31.80} & \multicolumn{1}{r}{48.86} \\
q$_{prev}$ old model     & \multicolumn{1}{r}{7.12}  & \multicolumn{1}{r}{7.92}  & \multicolumn{1}{r}{9.44}  &  & \multicolumn{1}{r}{16.89} & \multicolumn{1}{r}{20.67} & \multicolumn{1}{r}{26.08} &  & \multicolumn{1}{r}{6.33}  & \multicolumn{1}{r}{6.41}  & \multicolumn{1}{r}{6.50}  &  & \multicolumn{1}{r}{21.55} & \multicolumn{1}{r}{29.96} & \multicolumn{1}{r}{46.00} \\
q$_{prev}$ new model     & \multicolumn{1}{r}{46.27} & \multicolumn{1}{r}{58.31} & \multicolumn{1}{r}{68.13} &  & \multicolumn{1}{r}{3.10}  & \multicolumn{1}{r}{3.92}  & \multicolumn{1}{r}{7.36}  &  & \multicolumn{1}{r}{72.07} & \multicolumn{1}{r}{74.29} & \multicolumn{1}{r}{75.87} &  & \multicolumn{1}{r}{21.91} & \multicolumn{1}{r}{25.69} & \multicolumn{1}{r}{52.01} \\ 
\multicolumn{1}{l}{}  &                           &                           &                           &                           &                           &                           &                           &                           &                           &                           &                           &                           \\\hline
\end{tabular}
\end{table*}

The last paper that described the past and future motion of LPCs under simultaneous Galactic and stellar perturbations \citep{kroli_dyb:2017} used a list of 90 potential stellar perturbers selected on the basis of data from the XHIP catalogue \citep{anderson_francis:2011} which was a compilation of the Hipparcos mission results augmented with radial velocities available at that time. In that paper \cite{kroli_dyb:2017} studied a hundred of LPCs with large osculating perihelion distances and found no strong stellar perturbation effects.

Now we offer the updated list of 643 potential perturbers. Some of them can perturb cometary motion in a spectacular way, see \cite{First-stars:2020}. But the effect of the newly obtained list of perturbers on motion of many more comets is also noticeable and very important. A paper describing this in great detail is in preparation, here we only show selected results for four comets to serve as an example. In Table~\ref{tab:q-prev} we present previous perihelion distances (one orbital period to the past) for C/1993 F1, C/1997 BA6, C/1999 N4, and C/2006 E1, obtained using three different dynamical models. The starting perihelion distances of the original orbits are presented in the first row. Next three rows consists of previous perihelion distance values obtained without stellar perturbations, with an old dynamical model used in \cite{kroli_dyb:2017} and using the list of stellar perturbers announced in this paper, respectively.  These values are presented here as three deciles (10-50-90 percent) since the distribution of clones used for the uncertainty estimation is far from being Gaussian.

In \cite{kroli_dyb:2017} a comet is classified as dynamically old if its previous perihelion distance is below 10~au. On the other hand, it is called dynamically new if the previous perihelion distance appears to be greater than 20~au. As it is easy to note, the usage of the new list of stellar perturbers reversed this classification for three of the presented comets. Such a change is noticed in a large percentage of almost 300 studied LPCs and will be described in a future paper (Dybczyński \& Królikowska, in preparation). 

But there exist one important problem with the list of stellar perturbers announced in this paper -- its name is ALS\,9243. As it was described in detail in Sect.~\ref{als9243} there exist a fundamental inconsistency in data available for this star: its distance obtained with the \textit{Gaia} mission strongly disagrees  with its previously published spectral type and luminosity class. This controversy may lead to two solutions:

\begin{itemize}
\item[\textbullet] The distance based on \textit{Gaia} DR2 is completely wrong, the star is much farther (say, over 2~kpc) and this completely rules out this star from a list of potential perturbes.
\item[\textbullet] The astrometric results presented in \textit{Gaia} DR2 are approximately correct, which makes this star an important perturber since its nominal minimal distance from the Sun is as small as 0.25\,pc and the approach occurred over 2 Myr ago, so almost all the observed LPCs can be affected it. But, to be able to calculate this  effect, we need the mass of this star.
\end{itemize}

As it was described in detail in Sect.~\ref{als9243}, the most often quoted spectral type and luminosity class of ALS\,9243 is O9 IV (see Table~\ref{tab:spectra}) which results in the mass of 18.5 solar masses. Since these astrophysical parameters are in contradiction with the \textit{Gaia} DR2 astrometry and such a close passage of a hot and massive star 2~Myr ago seems improbable, we have to make a trade-off. For now, waiting for new data, we decided to use the \textit{Gaia} DR2 astrometry but we assume the mass of ALS\,9243 to be equal to 2.0 solar masses.

The correct mass value of ALS\,9243 might be crucial for the LPCs dynamical history studies as we show in Table~\ref{tab:q-masy}. Here we used nominal orbits of the same comets as shown in Table~\ref{tab:q-prev} however, we calculate their previous perihelion distance assuming different masses for ALS\,9243. 
In the first row, labelled 'O', the starting value of the perihelion distance is shown. In the next six rows the previous perihelion distance of these four comets are presented, calculated using the assumed mass of ALS\,9243 shown in the first column. For the sake of additional comparison, in the last row, labelled 'G', we present the previous perihelion distance obtained without stellar perturbations at all, only as a result of the galactic perturbations. 

One can easily observe that comets C/1993~F1 and C/1999~N4 are practically not affected by ALS\,9243 at all. But the remaining two comets are affected significantly by this star and their previous perihelion distance depends strongly on its assumed mass.

\begin{table}
    \centering
    \caption{The influence of the assumed ALS\,9243 mass on the previous perihelion distance of the same comets as presented in Table~\ref{tab:q-prev}. The first row, labelled 'O', presents the starting perihelion distance of the original orbit. The last row, labelled 'G', presents the previous perihelion distance when all stellar perturbations were omitted. In the remaining rows the first column shows the assumed mass of ALS\,9243 expressed in solar masses.} 
    \label{tab:q-masy}
    \begin{tabular}{r r r r r} 
    \hline
    &&&&\\
     & C/1993 F1 & C/1997 BA6 & C/1999 N4 & C/2006 E1 \\
     &&&&\\
    \hline
    &&&&\\
O:  &     5.8995 & 3.4404 & 5.5139 & 6.0361\\
18.5  &    58.0720 & 1486.4459 & 73.7724 & 2513.8460\\
10.0  &    58.0730 & 361.4909 & 73.7736 & 656.4899\\
5.0  &    58.0735 & 60.0877 & 73.7743 & 137.7216\\
2.0  &    58.0738 & 3.7164 & 73.7747 & 22.0914\\
1.0  &    58.0740 & 5.5040 & 73.7748 & 15.6289\\
0.0  &    58.0741 & 17.6174 & 73.7749 & 25.1433\\
G:  &     8.1556 & 19.6341 & 6.4411 & 31.8632\\
&&&&\\
         \hline
    \end{tabular}
\end{table}

\section{Conclusions and prospects}
\label{conclusions}

Due to a great increase of our knowledge on the Galactic neighbourhood of the Sun we were able to significantly correct and update the list of potential stellar perturbers of long period comets motion. The full list of analysed objects includes 751 single stars and 69 stellar systems. Among them 643 objects appeared to have their minimal heliocentric distance smaller than 4~pc and therefore are classified by us as potential perturbers of LPCs motion.

Our updated list consists both of new stars or stellar systems found by us in a manner described in Sect.~\ref{compiling_list} and of stars that were previously classified as stellar perturbers in the earlier papers mentioned in Sect.~\ref{compiling_list}. Objects from the later group were thoroughly examined whether new or more accurate data are available. Thanks to the improvement in the quality of data we were able to verify the importance of each perturber, how it changes due to new measurements, and, more importantly, check whether the star in question is a component of a multiple system. In some cases taking the multiplicity into account resulted in removing the perturber from the list, in other cases, this approach just changed the expected value of the minimal heliocentric distance and therefore the potential influence of this particular perturber on LPCs motion. 

The examples presented in Sect.~\ref{multiple_examples} show the importance of taking the multiplicity into account  and many drawbacks of still incomplete data which sometimes limited and hindered us from calculating center of mass parameters based on data concerning all known components of the considered system. Main issues are: lack of radial velocities measurements and masses (which is also applicable to single stars), different epochs of measurements of the positions and proper motions of components of the systems which can lead to unrealistic  results, and incomplete data on multiplicity of systems, especially these with stars for the first time observed by \textit{Gaia} mission. These issues may be addressed with the future \textit{Gaia} data releases.  

During gathering of data necessary to calculate minimal heliocentric distances of the stars, we came across a star designated as ALS\,9243 which, basing on available measurements (from \textit{Gaia} DR2 catalogue augmented with radial velocity from SIMBAD database), 2.5 Myr ago visited the vicinity of the Sun at the distance equal to 0.25 pc. This object has never been mentioned as a stellar perturber before and before the \textit{Gaia} mission its distance was often estimated to be more than 1~kpc. Because of the discrepancies in mass estimates, we decided to check the atmospheric parameters of the star in question. Our results indicates that this star seems to be very massive, even up to 22 $M_\odot$ which is inconsistent with its distance presented in \textit{Gaia} DR2. We hope that, in the nearest future, we will be given an opportunity to obtain consistent data and verify the significance of this perturber. Before the clarification of this puzzling case, we decided to use \textit{ Gaia} DR2 astrometry and adopt 2.0~M$_{\odot}$ as the mass of ALS\,9243 in our database. 

All data used by us (with sources) and our results are gathered in a small publicly available database of potential perturbers. This might be a useful tool in future dynamical studies of near-parabolic comets dynamics.

\begin{acknowledgements}
We would like to thank the anonymous referee for constructive suggestions that greatly improved the quality of this manuscript. This research was partially supported by the project 2015/17/B/ST9/01790 founded by the National Science Center in Poland. MP wants to thank Ewa Niemczura from the Astronomical Institute of the University of Wrocław, for a valuable advice. PAD would like to thank Marcin Gawroński from the Astronomical Institute of the Nicolaus Copernicus University in Toruń for his valuable comments on the nature of ALS\,9243. We would like to thank Kiril Stoyanov and Dragomir Marchev who provided us with spectra of ALS\,9243 and Wojciech Dimitrow for insightful comments on the interpretation of these spectra. This research has made use of the SIMBAD database, operated at CDS, Strasbourg, France. This research has also made use of the VizieR catalogue access tool, CDS, Strasbourg, France (DOI: 10.26093/cds/vizier). This work has made use of data from the European Space Agency (ESA) mission {\it Gaia} (\url{https://www.cosmos.esa.int/gaia}), processed by the {\it Gaia} Data Processing and Analysis Consortium (DPAC,
\url{https://www.cosmos.esa.int/web/gaia/dpac/consortium}). Funding for the DPAC
has been provided by national institutions, in particular the institutions
participating in the {\it Gaia} Multilateral Agreement.
\end{acknowledgements}

\bibliographystyle{aa.bst}
\bibliography{PAD30}

\end{document}